\documentclass[aps,pra,twocolumn,showpacs]{revtex4}

\usepackage{bm}
\usepackage{epsfig}

\def\nb #1{{\hbox{\bf #1}}}

\newcommand{\gsim}{\buildrel{>}\over{\sim}}
\newcommand{\lm}{{\cal P}}

\begin{document}

\title{Nonlinear waves in a cylindrical Bose-Einstein condensate}
\author{S. Komineas$^1$ and N. Papanicolaou$^2$}
\affiliation{$^1$Physikalisches Institut, Universit\"at Bayreuth, 
D-95440 Bayreuth, Germany \\
$^2$Department of Physics, University of Crete, and Research Center of Crete,
Heraklion, Greece}

\date{\today}

\begin{abstract}
We present a complete calculation of solitary waves propagating in a
steady state with constant velocity $v$ along a cigar-shaped 
Bose-Einstein trap approximated as infinitely-long cylindrical.
For sufficiently weak couplings (densities) the main features
of the calculated solitons could be captured by effective one-dimensional (1D)
models. However, for stronger couplings of practical interest, the
relevant solitary waves are found to be hybrids of quasi-1D solitons
and 3D vortex rings. An interesting hierarchy of vortex rings
occurs as the effective coupling constant is increased through
a sequence of critical values.
The energy-momentum dispersion of the above structures is shown to
exhibit characteristics similar to a mode proposed sometime ago by Lieb 
within a strictly 1D model, as well as some rotonlike features.
\end{abstract}

\maketitle

\section{Introduction}
Solitary waves that may occur in a Bose-Einstein Condensate (BEC) have been
traditionally discussed in terms of the classical Gross-Pitaevskii (GP) 
model which is appropriate for the description of weakly correlated systems
\cite{dalfovo}. For instance, a simple soliton was obtained by
Tsuzuki \cite{tsuzuki} in a homogeneous 1D model, while Zakharov and
Shabat \cite{zakharov} developed inverse-scattering techniques for the
study of multisolitons. Interestingly, the elementary soliton
proved to be relevant for an accurate semiclassical description
\cite{kulish,ishikawa} of an intriguing mode proposed earlier
by Lieb \cite{lieb} in a full quantum treatment of a 1D Bose gas
based on the Bethe Ansatz \cite{lieb2}.

The above developments had long remained purely theoretical
because of the absence of a physical realization of a strictly 1D
Bose gas. Nevertheless, the picture has significantly changed with
the recent observation of similar coherent structures in
confined BECs of alkali-metal atoms \cite{burger,denschlag}.
The very method of experimental production of solitary waves
(phase imprinting) was inspired by the analytical structure of the
1D soliton, while various effective 1D models have been developed for their
theoretical investigation
\cite{perez,jackson,muryshev,feder,dikande,muryshev2,salasnich}.
On the other hand, the actual stability of the theoretically predicted
1D solitary waves should be questioned within the proper 3D
environment of realistic traps \cite{muryshev,feder,dikande}.
An important step in that direction was the experimental observation
\cite{anderson} that a dark soliton initially created in a finite
trap eventually decays into vortex rings, as is also predicted by a
numerical solution of the corresponding initial-value
problem in a 3D classical GP model \cite{feder}.

Therefore, it is important to carry out a calculation of potential
nonlinear modes without {\it a priori} assumptions about their effective
dimensionality. One could envisage a picture in which the actual
solitary waves are hybrids of quasi-1D solitons and 3D vortex rings. It is
the aim of the present paper to make the above claim precise by calculating
solitary waves that propagate along a cylindrical trap in a steady state
with constant velocity $v$. Our approach was motivated by the calculation 
of vortex rings in a homogeneous BEC due to Jones and Roberts \cite{jones1}
and a similar calculation of semitopological solitons in planar ferromagnets
\cite{semi}.

We have already described the main result of this work in a recent short
communication \cite{komineas} but a substantial elaboration is
necessary in order to appreciate its full significance.
Thus the problem is formulated in Sec.~II where we also present
a brief but complete recalculation of the ground state and the
corresponding  linear (Bogoliubov) modes for comparison. A detailed
calculation of nonlinear modes is given in Sec.~III and the main
conclusions are summarized in Sec.~IV.

\section{Formulation and linear modes}
The physical picture that we have in mind is a slight idealization of the
experiment in Ref. \cite{burger}. Thus we consider a cigar-shaped  trap
filled with atoms of mass $m$. The transverse confinement frequency is
denoted by $\omega_\bot$ and the corresponding oscillator length
by $a_\bot \!=\!(\hbar/m\omega_\bot)^{1/2}$. The longitudinal
confinement frequency $\omega_\|$ is assumed to be much smaller
than $\omega_\bot$, hence we make the approximation of an
infinitely-long cylindrical trap with $\omega_\| \!=\! 0$.
Accordingly, complete specification of the system requires as input
the average linear density $\nu$ which is the number of atoms per
unit length of the cylindrical trap. Finally, we consider the two
dimensionless combinations of parameters:
\begin{equation}  \label{eq:21}
  \gamma=\nu a\,, \quad \gamma_\bot = \nu a_\bot\,,
\end{equation}
where $a$ is the scattering length related to the coupling constant
as usual by $U_0\!=\!4\pi\hbar^2 a/m$.

Now, in the actual experiment of Ref. \cite{burger}, the trap is
filled with $^{87}$Rb atoms,
the transverse frequency is chosen as 
$\omega_\bot \!=\! 2\pi \times 425$ Hz, the oscillator length
is calculated to be $a_\bot \!\approx\! 0.5\mu m$, and the estimated
linear density lies in the range $\nu \! > 0.1\, \hbox{atoms}/\hbox{\AA}$.
Also taking into account a scattering length $a\!\approx\! 50\,\hbox{\AA}$,
the dimensionless parameters (\ref{eq:21}) take values in
$\gamma\!>\! 5$ and $\gamma_\bot\!>\!5\times 10^2$. In fact,
our subsequent calculations will be carried out for a much wider range
of the above parameters. Therefore, apart from the idealization of a
cylindrical trap, our results are fairly realistic and could be
applied to a number of cases of experimental interest.

It is useful to introduce rationalized units through the rescalings
\begin{equation}  \label{eq:22}
t\!\to\! \frac{t}{\omega_\bot},\quad
{\nb r}\!\to\! a_\bot {\nb r},\quad
\Psi\!\to\! \frac{\nu^{1/2}}{a_\bot}\, \Psi.
\end{equation}
The energy functional extended to include a chemical potential is
then given by
\begin{equation}  \label{eq:23}
  W = \frac{1}{2} \int{\left[ \bm{\nabla}\Psi^* \bm{\nabla}\Psi
 + \rho^2\,\Psi^*\Psi + g (\Psi^*\Psi)^2 - 2\mu\,\Psi^*\Psi \right]\, dV},
\end{equation}
where $g\!=\!4\pi\gamma$ and $\rho^2\!=\!x^2+y^2$. Eq.~(\ref{eq:23})
yields energy $W$ in units of $\gamma_\bot (\hbar\omega_\bot)$ whereas
the chemical potential $\mu$ is measured in units of $\hbar \omega_\bot$.
The corresponding rationalized equation of motion reads
\begin{equation}  \label{eq:24}
  i \frac{\partial \Psi}{\partial t} = -\frac{1}{2} \Delta\Psi
+ \frac{1}{2} \rho^2\Psi + g (\Psi^*\Psi) \Psi - \mu \Psi\,,
\end{equation}
and depends only on the dimensionless coupling constant $\gamma$,
because $g\!=\!4\pi\gamma$ and the chemical potential
$\mu\!=\!\mu(\gamma)$ is fixed by the requirement that the system
carry in its ground state a definite average linear density $\nu$.

\begin{figure}
   \begin{center}
   \epsfig{file=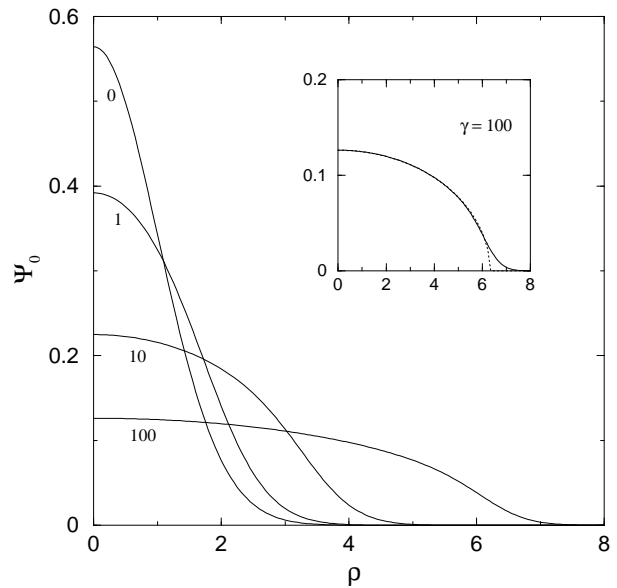,width=8cm}
   \end{center}
   \caption{Radial dependence of the ground-state wave function for
four representative values of the dimensionless coupling constant
$\gamma\!=\!0,1,10$ and 100. The corresponding values of the chemical
potential were calculated to be $\mu\!=\!1,\, 2.2571,\, 6.4324$, and 20.0431,
in units of $\hbar \omega_\bot$. The inset compares the accurate
numerical solution (solid line) with the TF approximation (\ref{eq:27})
(dotted line) for the strong coupling $\gamma\!=\! 100$.
Distance is measured in units of $a_\bot$.
   }
   \label{fig:1}
\end{figure} 

An important first step is thus to obtain accurate information about the
ground-state wave function $\Psi\!=\!\Psi_0(\rho)$ which is normalized
according to
\begin{equation}  \label{eq:25}
 \int_0^\infty{2\pi\rho\,d\rho\, |\Psi_0|^2} = 1\,,
\end{equation}
to conform with our choice of rationalized units. The wave function
$\Psi_0(\rho)$ is numerically calculated as the minimum of the energy
functional, under the constraint (\ref{eq:25}) that
fixes the chemical potential $\mu(\gamma)$, by a variant of a relaxation
algorithm \cite{dalfovo2}.  Explicit results are illustrated in
Fig.~\ref{fig:1} for some typical values of $\gamma$ where we also
quote the corresponding values of the chemical potential.

The preceding numerical determination of the ground state will provide the
basis for all subsequent calculations. However, it is worth mentioning
here some limiting cases where the ground state is known analytically.
At $\gamma\!=\!0$,
\begin{equation}  \label{eq:26}
  \Psi_0 = \frac{1}{\sqrt{\pi}}\, e^{-\rho^2/2}\,,
\end{equation}
and the chemical potential degenerates to $\mu\!=\!1$. In the opposite limit,
$\gamma\!\gg\! 1$, one may use the Thomas-Fermi (TF) approximation
\cite{baym}
\begin{equation}  \label{eq:27}
  \Psi_0 = \left[ \frac{2}{\pi R_\bot^2} \left(1-\frac{\rho^2}{R_\bot^2}
       \right) \right]^{1/2}
\end{equation}
for $0\!\leq\!\rho\!\leq\!R_\bot\!=\!2\gamma^{1/4}$, and $\Psi_0\!=\! 0$
for $\rho\!>\!R_\bot$. The chemical potential is given accordingly by
$\mu\!\approx\!2\,\gamma^{1/2}$. A comparison with the accurate numerical
solution is shown in the inset of Fig.~\ref{fig:1} for $\gamma\!=\! 100$.
In fact, as we shall see shortly, the TF approximation provides a
reasonable description of some quantities of physical interest even
for $\gamma\!\sim\! 1$.

We will also need some information from the linear (Bogoliubov) modes
which have already been calculated in the literature to varying degree
of completeness \cite{zaremba,stringari,fedichev}.
Here we employ a numerical algorithm of our own briefly described as follows.
Equation (\ref{eq:24}) is linearized by inserting
$\Psi\!=\!\Psi_0 + f + i\chi$ where $\Psi_0(\rho)$ is the calculated
ground-state wave function while $f\!=\!f({\nb r},t)$ and
$\chi\!=\!\chi({\nb r},t)$ are real functions that account for small
fluctuations around the ground state. It is somewhat more convenient to use
the linear combinations $a\!=\!f+\chi$ and $b\!=\!f-\chi$ which satisfy
the linearized equations
\begin{equation}  \label{eq:28}
\frac{\partial}{\partial t} \left( \begin{array}{c} a \\ b \end{array} \right)
 = M \left( \begin{array}{c} a \\ b \end{array} \right)\,, \quad
M \equiv \left( \begin{array}{cc} -V & -D \\ D & V \end{array} \right)
\end{equation}
where
\begin{equation}  \label{eq:29}
V = g \Psi_0^2,\quad D = -\frac{1}{2} \Delta + \frac{1}{2} \rho^2+2\, V-\mu\,.
\end{equation}
Our task is then to calculate the spectrum of the differential operator $M$ 
whose eigenvalues are purely imaginary and come in pairs $\pm i \omega$
where $\omega$ is the sought after physical frequency.

We restrict attention to axially-symmetric waves that propagate
along the $z$ axis with wave number $q$. The Laplace operator is then
replaced by
\begin{equation}  \label{eq:210}
\Delta = \frac{\partial^2}{\partial \rho^2} 
       + \frac{1}{\rho}\frac{\partial}{\partial \rho} - q^2
\end{equation}
and the amplitudes $a$ and $b$ may be assumed to depend only on the radial
distance $\rho$. A finite-matrix approximation of the operator
$M$ is obtained by expanding both $a$ and $b$ in terms of a basis set
of non-orthogonal Gaussian wave packets with randomly chosen oscillator 
lengths \cite{papspathis}. It is also prudent to enlarge the basis
set by including the ground-state wave function $\Psi_0(\rho)$ itself,
in order to directly account for the zero (Goldstone) mode associated with the
number symmetry. The resulting algorithm is then quite efficient and provides
stable approximations of the low-lying eigenvalues even if we include 
a small number of basis elements.

\begin{figure}
   \begin{center}
   \epsfig{file=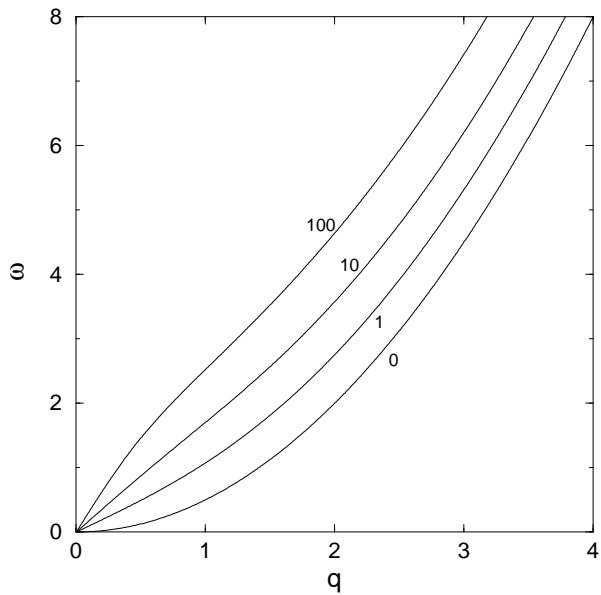,width=8cm}
   \end{center}
   \caption{The lowest branch in the Bogoliubov spectrum for four
representative values of the dimensionless coupling constant
$\gamma\!=\!0,1,10$ and 100. The frequency $\omega$ is measured 
in units of $\omega_\bot$ and the wave number $q$ in units of $1/a_\bot$.
The corresponding values of the speed of sound were calculated to be
$c\!=\!0,\, 0.95,\, 1.77$, and 3.17, in units of $a_\bot \omega_\bot$.
   }
   \label{fig:2}
\end{figure}

In Fig.~\ref{fig:2} we present explicit results for the lowest 
eigenfrequency $\omega\!=\!\omega(q)$ for the same set of coupling constants
as in Fig.~\ref{fig:1}. At $\gamma\!=\!0$, $\omega(q)$ reduces to the
free-particle quadratic dispersion $\omega\!=\!q^2/2$, as expected.
At nonzero $\gamma$, the dispersion becomes linear near the origin,
$\omega\!\approx\!c|q|$, where $c$ is the speed of sound for
which explicit values are also quoted in Fig.~\ref{fig:2}.
Finally, we note that our results are in apparent agreement with the
Bogoliubov dispersion calculated earlier within the TF approximation
\cite{zaremba,stringari} as well as numerically \cite{fedichev} -- even
though a different parameterization of the spectrum was employed 
in the latter reference.

\begin{figure}
   \begin{center}
   \epsfig{file=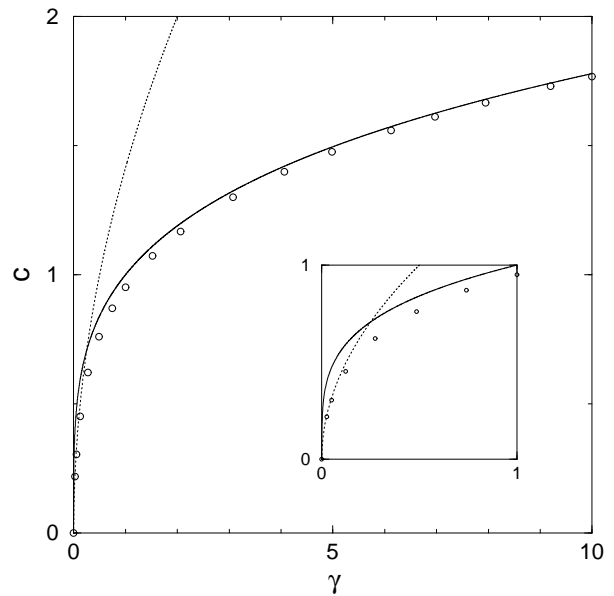,width=8cm}
   \end{center}
   \caption{Speed of sound $c$ in units of $a_\bot \omega_\bot$ as a function
of the dimensionless coupling constant $\gamma$. Open circles stand
for our numerical data, the solid line for the TF asymptote of 
Eq.~(\ref{eq:211}), and the dotted line for the weak-coupling asymptote of
Eq.~(\ref{eq:212}).
   }
   \label{fig:3}
\end{figure}

The speed of sound is a quantity of special physical interest and will
also play an important role in the theoretical development of Sec.~III.
Hence we have carried out a calculation for a wider set of coupling constants
and the results are summarized in Fig.~\ref{fig:3}.
It is interesting that our accurate numerical results are consistent
with the TF approximation \cite{zaremba,stringari,kavoulakis}
\begin{equation}  \label{eq:211}
 c \approx  \gamma^{1/4}
\end{equation}
even for values of $\gamma$ as low as 1, where the error is about 5\%,
whereas the error is reduced to less than 1\% for $\gamma\!\gsim\! 10$. 
This fact is especially important because Eq.~(\ref{eq:211}) was employed for
the analysis of experimental data \cite{andrews}.
The relative accuracy of this approximation progressively deteriorates
in the region $\gamma\!<\! 1$, but a new asymptote, namely
\begin{equation}  \label{eq:212}
 c \approx (2 \gamma)^{1/2}\,,
\end{equation}
was predicted to be reached for sufficiently weak couplings \cite{jackson}.
The weak-coupling approximation (\ref{eq:212}) is actually consistent
with our numerical data for $\gamma\!<\! 1/4$, as is shown in the
inset of Fig.~\ref{fig:3}. However, we should add that the linear part 
of the Bogoliubov dispersion becomes very narrow in this region of couplings.

\section{Nonlinear waves}
We now turn to the calculation of axially-symmetric solitary waves traveling along the $z$ axis in a steady state with constant velocity $v$.
These are described by a wave function of the form $\Psi\!=\!\Psi(\rho,\xi)$,
with $\xi\!=\!z-v t$, which is inserted in Eq.~(\ref{eq:24}) to yield
the stationary differential equation
\begin{eqnarray}
\label{eq:31}
 - i v\, \frac{\partial\Psi}{\partial\xi} & = & -\frac{1}{2} \Delta\Psi
 + \frac{1}{2} \rho^2\Psi + g (\Psi^*\Psi) \Psi - \mu \Psi\,, \\
\noalign{\medskip}
 \Delta & = & \frac{\partial^2}{\partial\rho^2} + \frac{1}{\rho}
 \frac{\partial}{\partial\rho} + \frac{\partial^2}{\partial\xi^2}\,.
\nonumber
\end{eqnarray}
The wave function must vanish in the limit $\rho\to \infty$, thanks to the
transverse confinement, while the condition
\begin{equation}  \label{eq:32}
  \lim_{\xi\to \pm \infty}  |\Psi(\rho,\xi)| = |\Psi_0(\rho)|
\end{equation}
enforces the requirement that the local particle density coincide asymptotically
with that of the ground state calculated in Sec.~II.
But the phase of the wave function is not fixed
{\it a priori} at spatial infinity except for a mild restriction
implied by the von Neumann boundary condition
\begin{equation}  \label{eq:33}
  \lim_{\xi\to \pm \infty} \frac{\partial \Psi}{\partial \xi} = 0
\end{equation}
adopted in our numerical calculation. Our task is then to find
concrete solutions of Eq.~(\ref{eq:31}) that satisfy the
boundary conditions just described.

An important check of the numerical calculation is provided by the
virial relation
\begin{equation}
\label{eq:34}
v\, \lm = \int{\left[\frac{1}{2} \frac{\partial\Psi^*}{\partial\xi}
\frac{\partial\Psi}{\partial\xi} + \rho^2\,\Psi^*\Psi
+ \frac{g}{2} (\Psi^*\Psi)^2 - \mu\, \Psi^*\Psi \right] dV}\,,
\end{equation}
obtained by standard scaling arguments \cite{semi}.
Here $\lm$ is the linear momentum given by the usual definition
\begin{equation}
\label{eq:35}
\lm = \frac{1}{2 i}\, \int{\left(\Psi^* \frac{\partial\Psi}{\partial z}
-\frac{\partial\Psi^*}{\partial z}\Psi\right)\; dV} =
\int{n\, \frac{\partial\phi}{\partial z}\; dV},
\end{equation}
and is measured in units of $\hbar \nu\!=\!\gamma_\bot (\hbar/a_\bot)$.
In the second step of Eq.~(\ref{eq:35}) we employ
hydrodynamic variables defined from 
\begin{equation}  \label{eq:36}
\Psi=\sqrt{n}\, e^{ i\phi}\,,
\end{equation}
where $n\!=\!|\Psi|^2$ is the local particle density and the phase $\phi$ 
may be used to construct the velocity field ${\nb u} = \bm{\nabla}\phi$.

Numerical solutions of Eq.~(\ref{eq:31}) are obtained by an iterative
Newton-Raphson algorithm \cite{jones1,semi} briefly described
as follows. Suppose that $\Psi\!=\!\Psi_{\rm in}$ is an initial rough
guess for the solution at some velocity $v$. We then insert in
Eq.~(\ref{eq:31}) the configuration $\Psi_{\rm out}\!=\!\Psi_{\rm in}+X$
and keep terms that are at most linear in the amplitude $X$.
Thus we derive an inhomogeneous differential equation of the form
$L X \!=\! Y$ where the linear operator $L$ and the source $Y$ are both
calculated in terms of $\Psi_{\rm in}$. We solve this linear system
for $X\!=\! L^{-1} Y$ to obtain $\Psi_{\rm out}\!=\!\Psi_{\rm in}+L^{-1} Y$
which is used as input for the next iteration until convergence is
achieved at some specified level of accuracy.
The procedure is repeated by incrementing the velocity to a different value,
typically in steps of $\delta v \!=\! \pm 0.01$, using as input
the converged configuration obtained at the preceding value of the velocity.
Therefore, the main numerical burden consists of constructing a
finite-matrix lattice approximation of the linear operator $L$ which is then inverted by standard routines appropriate  for sparse linear systems.

The Newton-Raphson algorithm typically converges  after a few iterations and
the final configuration is independent of the specific choice $\Psi_{\rm in}$.
But it is also clear that the algorithm will not converge for most
choices of $\Psi_{\rm in}$.
Hence it is important to invoke an educated guess for the input configuration
provided by the product Ansatz:
\begin{equation}  \label{eq:37}
  \Psi_{\rm in} = [c_1 - i\, c_2 \tanh(c_3 \xi)]\, \Psi_0(\rho)
\end{equation}
which capitalizes on the analytically known solitary wave in the
homogeneous 1D model \cite{tsuzuki,zakharov} and the ground-state
configuration $\Psi_0(\rho)$ numerically calculated in Sec.~II.
The constants $c_1, c_2$ and $c_3$ are definite functions of the velocity
$v$ within the strictly 1D model, but such precise relations need
not be invoked for our current purposes except for the normalization
condition $c_1^2+c_2^2\!=\!1$ that is necessary to enforce the boundary 
condition (\ref{eq:32}). In other words, the above constants are
treated here as  trial parameters until we achieve convergence for a
specific velocity $v$.
A corrolary of the preceding discussion is that the converged configuration
does not depend on the precise choice of those parameters,
and it is certainly not in the form of a product  Ansatz often employed for
the derivation of effective 1D models \cite{jackson,salasnich}.
Finally, we note that the Ansatz (\ref{eq:37}) satisfies the parity
relations
\begin{equation}  \label{eq:38}
\hbox{Re}\,\Psi(\rho,\xi) = \hbox{Re}\,\Psi(\rho,-\xi), \quad
\hbox{Im}\,\Psi(\rho,\xi)  =  -\hbox{Im}\,\Psi(\rho,-\xi)
\end{equation}
which are compatible with Eq.~(\ref{eq:31}) and are actually satisfied
by all solutions constructed in the present paper.

\begin{figure}
   \begin{center}
   \epsfig{file=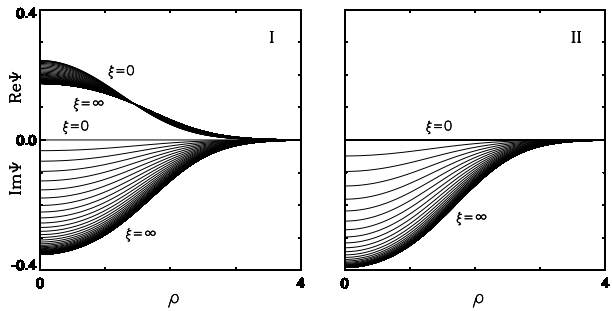,width=8cm,bbllx=40bp,bblly=450bp,bburx=545bp,bbury=700bp}
   \end{center}
   \caption{Solitary wave for $\gamma\!=\! 1$ and two representative values
of the velocity: $v\!=\!c/2$ (I) and $v\!=\! 0$ (II). We display
the radial dependence of the real and the imaginary part of the wave function
for various positive values of $\xi$ in steps of $\delta\xi\!=\!0.1$.
The corresponding results for negative $\xi$ are obtained through the 
parity relations (\ref{eq:38}).
   }
   \label{fig:4}
\end{figure}

We begin with the special case of the relatively weak coupling $\gamma\!=\!1$
for which the speed of sound was calculated to be $c\!=\!0.95$
in Sec.~II. The simplest possibility is to first attempt to derive
a static ($v\!=\!0$) soliton starting with the input configuration
(\ref{eq:37}) applied for, say, $c_1\!=\!0$ and $c_2\!=\!1\!=\!c_3$.
Indeed, the algorithm quickly converges to a wave function with a
nontrivial imaginary part but vanishing real part.
The velocity is then incremented to positive values in steps of
$\delta v\!=\!0.01$ and the corresponding wave functions acquire also 
a nontrivial real part. The process may be continued until the velocity 
$v$ approaches the speed of sound $c$ beyond which the solitary wave
ceases to exist. An equivalent sequence of solitary waves
with velocities in the range $-c\!<\!v\!\leq 0$ is obtained
either by starting again with the $v\!=\!0$ soliton and pushing it to
negative velocities or, simply, by taking the complex conjugate of the 
wave function calculated for $0\!\leq v \!<\! c$, since
\begin{equation}  \label{eq:39}
  v \to - v, \quad \Psi \to \Psi^*\,,
\end{equation}
is an obvious symmetry of Eq.~(\ref{eq:31}). A detailed illustration of the
calculated solitary wave function is given in Fig.~\ref{fig:4} for two
representative values of the velocity: $v\!=\!c/2$ and $v\!=\!0$.

\begin{figure}
   \begin{center}
   \epsfig{file=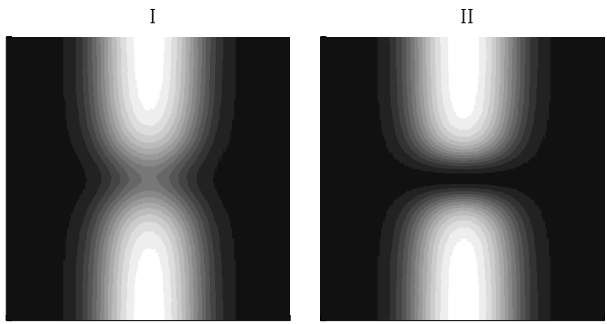,width=8cm,bbllx=25bp,bblly=460bp,bburx=560bp,bbury=745bp}
   \end{center}
   \caption{Contour levels of the local particle density
$n\!=\!|\Psi|^2$ for $\gamma\!=\! 1$, in a plane that contains the $z$ axis
and cuts across the cylindrical trap. The complete 3D picture may be
envisaged by simple revolution around the $z$ axis.
Regions with high particle density are bright while regions with
zero density are black. The two special cases considered are the same as in
Fig.~\ref{fig:4}.
   }
   \label{fig:5}
\end{figure}

A partial but more transparent illustration is given in Fig.~\ref{fig:5}
which depicts the level contours of the local particle density $n\!=\!|\Psi|^2$
for the two special cases considered in Fig.~\ref{fig:4}.
In words, the calculated solitary wave is a mild soundlike disturbance of the ground state when $|v|$ approaches the speed of sound $c$, while it becomes
an increasingly dark soliton with decreasing $|v|$ and reduces to 
a completely dark (black) soliton at $v\!=\!0$.

It is now important to calculate the energy-momentum dispersion of the
solitary wave. The excitation energy is defined as
\begin{equation}  \label{eq:310}
  E = W-W_0
\end{equation}
where both $W$ and $W_0$ are calculated from Eq.~(\ref{eq:23}) applied
for the solitary wave $\Psi(\rho,\xi)$ and the ground state
$\Psi_0(\rho)$, respectively. The presence of the chemical potential
in Eq.~(\ref{eq:23}) provides the compensation that is necessary to
compare energies of states with the same number of particles.
Similarly, the relevant physical momentum is not the linear momentum $\lm$
of Eq.~(\ref{eq:35}) but the impulse $Q$ defined in a manner analogous 
to the case of a homogeneous gas \cite{kulish,ishikawa},
\begin{eqnarray}
\label{eq:311}
 Q  & = & \int{(n-n_0)\, \frac{\partial\phi}{\partial z}\, dV}
          = \lm - \delta\phi\,, \\
\noalign{\medskip}
 \delta \phi & \equiv & \int_0^\infty{2\pi\rho\, d\rho\; n_0(\rho)
[\phi(\rho,z\!=\! \infty)-\phi(\rho,z\!=\! -\infty)]}\,, \nonumber
\end{eqnarray}
where $n_0\!=\!|\Psi_0(\rho)|^2$ is the ground-state particle density
and $\delta \phi$ is now the weighted average of the phase difference
between the two ends of the trap.
The delicate distinction between linear momentum and impulse
has been the subject of discussion in practically all treatments of
classical fluid dynamics \cite{batchelor,saffman} and continues to
play an important role in the dynamics of superfluids \cite{jones1}.
Here we simply postulate the validity of the definition of impulse in
Eq.~(\ref{eq:311}) and note that the corresponding group-velocity relation
\begin{equation}  \label{eq:312}
 v = \frac{d E}{d Q}
\end{equation}
is satisfied to an excellent accuracy in our numerical calculation
and thus provides a highly nontrivial check of consistency.
In turn, the virial relation (\ref{eq:34}) is verified using the 
standard definition of the linear momentum $\lm$ in Eq.~(\ref{eq:35}),
as expected. We finally note that the same phase difference
$\delta\phi$ which is important for experimental production
of solitary waves through phase imprinting \cite{burger,denschlag}
is also crucial for the calculation of the impulse.

\begin{figure}
   \begin{center}
   \epsfig{file=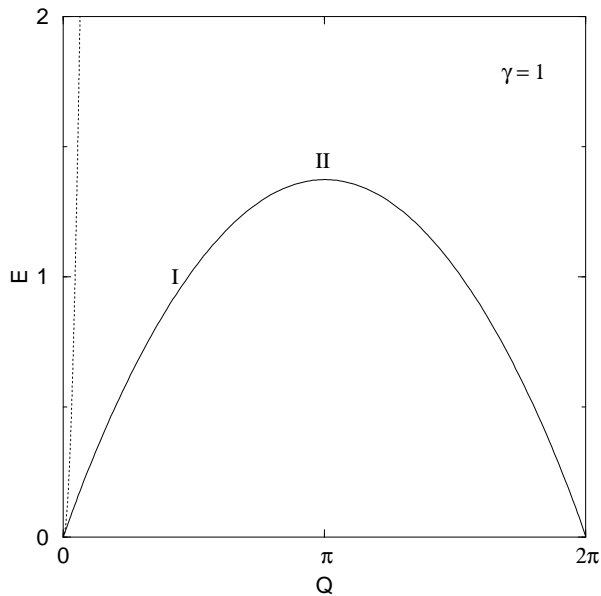,width=8cm}
   \end{center}
   \caption{Energy $E$ in units of $\gamma_\bot (\hbar\omega_\bot)$ versus
impulse $Q$ in units of $\gamma_\bot (\hbar/a_\bot)$ for $\gamma\!=\! 1$.
The solid line corresponds to the complete sequence of solitary waves 
discussed in the text, and the dotted line to the $\gamma\!=\! 1$ Bogoliubov
dispersion of Fig.~\ref{fig:2} adjusted to current units.
Symbols I and II correspond to the two special cases of the solitary wave
illustrated in Figs.~\ref{fig:4} and \ref{fig:5}.
   }
   \label{fig:6}
\end{figure}

The dispersion $E\!=\!E(Q)$ calculated for the complete sequence of
solitary waves with velocities in the range $-c\!<\!v\!<\!c$ is
illustrated in Fig.~\ref{fig:6}. The apparent $2\pi$ periodicity
seems surprising, but occured also in the original calculation
of a similar mode by Lieb \cite{lieb} within a full quantum treatment
of a 1D Bose gas interacting via a $\delta$-function potential.
The Lieb mode was later rederived by a fairly accurate 
semiclassical approximation based on the elementary solitary wave of the
1D classical GP model \cite{kulish,ishikawa}.

Lieb further argued that the corresponding Bogoliubov mode is no more
elementary and thus proposed an intriguing dual interpretation of the
excitation spectrum. It should be noted that the dispersions of the two
modes exhibit the same linear dependence at low momenta, $E\!\approx\!c|Q|$,
where $c$ is the speed of sound, but significant differences arise at
finite momenta. The differences are especially pronounced in the
current calculation within a cylindrical trap. Specifically, let us
assume an average linear density $\nu\!=\!0.02\, \hbox{atoms}/\hbox{\AA}$
which leads to $\gamma\!=\!\nu a\!=\! 1$ and 
$\gamma_\bot\!=\!\nu a_\bot\!=\! 10^2$. If we then adjust the $\gamma\!=\! 1$
Bogoliubov dispersion of Fig.~\ref{fig:2} to the units employed in
Fig.~\ref{fig:6}, the two dispersions are seen to diverge very quickly
at the scale of Fig.~\ref{fig:6}.
In other words, Bogoliubov and Lieb modes operate at rather different 
energy and momentum scales in a realistic trap.

To summarize the preceding accurate calculation for $\gamma\!=\!1$,
the solitary wave is essentially quasi-1D in this weak-coupling 
region and its main features are indeed captured by an effective 1D model
\cite{jackson2}.
However, all cases of actual experimental interest
\cite{burger,denschlag,anderson} are characterized by significantly
larger values of the effective coupling where quasi-1D solitons are
expected to be unstable \cite{muryshev,feder}.
In particular, Ref. \cite{muryshev} suggests that a critical coupling
occurs in a cylindrical trap when $n_{\rm max} U_0/\hbar\omega_\bot\!=\!2.4$,
where $n_{\rm max}$ is the maximum local particle density in the ground
state of the trap. If we tentatively assume that the TF approximation
(\ref{eq:27}) can be trusted at the anticipated critical coupling,
the above criticality condition reads $2\gamma^{1/2}\!=\!2.4$ where
$\gamma$ is the dimensionless effective coupling constant defined in
Eq.~(\ref{eq:21}). The root of this equation $\gamma\!=\!\gamma_c\!=\!1.44$
provides a critical coupling $\gamma_c$ above which quasi-1D solitons
are predicted to be unstable \cite{jackson2}.

In fact, our numerical calculation suggests that the quasi-1D nature of the
solitary wave is completely lost at a higher critical coupling, namely for
$\gamma\!>\!\gamma_1\!\approx\!3.9$. The emerging new picture is clear 
at $\gamma\!=\!10$ which is the special case described in our
recent short communication \cite{komineas}.
This case is reanalyzed and further extended in the continuation of the
present paper.

\begin{figure}
   \begin{center}
   \epsfig{file=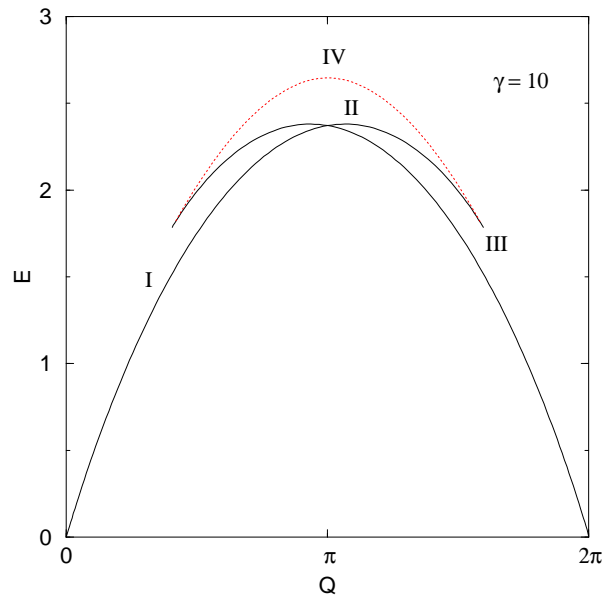,width=8cm}
   \end{center}
   \caption{Energy $E$ in units of $\gamma_\bot (\hbar\omega_\bot)$ versus
impulse $Q$ in units of $\gamma_\bot (\hbar/a_\bot)$ for $\gamma\!=\! 10$.
The solid line corresponds to the fundamental sequence of vortex rings
discussed in the text, and the dotted line to the auxiliary sequence that
contains the black soliton (point IV).
   }
   \label{fig:7}
\end{figure}

It is natural to begin again with the calculation of a static ($v\!=\!0$)
soliton obtained by using the input configuration (\ref{eq:37})
with $c_1\!=\!0, c_2\!=\! 1$ and practically any $c_3$.
We then increment the velocity to both positive and negative values in
steps of $\delta v\!=\!\pm 0.01$ to yield a sequence of solitary waves
which now display two surprising features. First, a ringlike
structure develops for $\gamma\!=\!10$ that was not present at $\gamma\!=\! 1$. Second, the above sequence exists only over the limited velocity
range $-v_1\!<\! v\!<\! v_1$ where $v_1\!=\!0.84\!=\!0.47\, c$
and $c\!=\!1.77$ is the speed of sound 
calculated in Sec.~II for $\gamma\!=\! 10$.
The existence of a critical velocity $v_1$ also becomes apparent in the
energy-momentum dispersion of the above sequence depicted by a dotted line
in Fig.~\ref{fig:7}. This portion of the dispersion is symmetric around
$Q\!=\!\pi$, where it achieves a maximum, but remains open ended
at two critical points that correspond to $v\!=\!\pm v_1$.

It is thus not surprising that an independent sequence of solitary waves with
lower energy exists for $\gamma\!=\! 10$.
Indeed, we return to the input configuration (\ref{eq:37}) but now target
a solution with velocity in the range $v_1\!<\! v\!<\! c$.
After some experimentation a solution is obtained for, say, $v\!=\!1.5$
if we choose the trial parameters $c_1\!=\!0.2$, $c_2\!=\!0.98$
and $c_3\!=\!3$. Having thus obtained a specific solution for
$v\!=\!1.5$ the algorithm is iterated forward and backwards in steps
of $\delta v\!=\! \pm 0.01$ to obtain an entirely new sequence of
solitary waves in the velocity range $-v_1\!<\! v\!<\! c$,
and a corresponding sequence for $-c\!<\! v\!<\!v_1$ through the
symmetry relation (\ref{eq:39}). Here $v_1$ is the same critical velocity
encountered in the preceding paragraph, as is also apparent in the
calculated energy-momentum dispersions which are depicted by solid lines
in Fig.~\ref{fig:7} and join the previously calculated dotted line
through cusps that correspond to $v\!=\!\pm v_1$.

\begin{figure}
   \begin{center}
   \epsfig{file=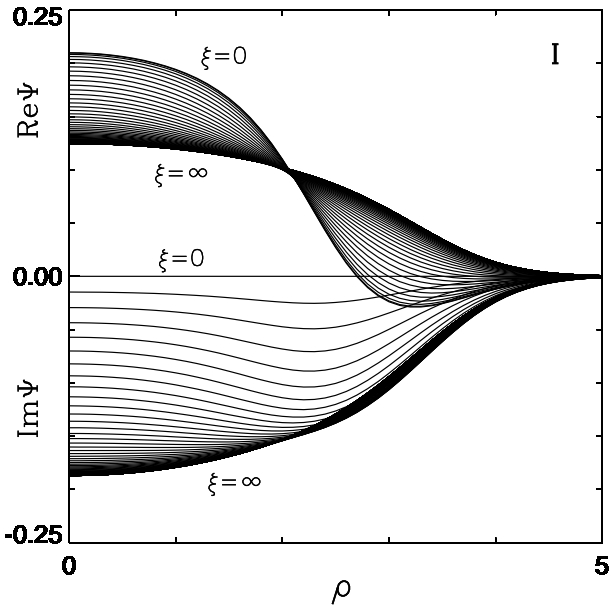,width=8cm,bbllx=140bp,bblly=435bp,bburx=425bp,bbury=720bp}
   \end{center}
   \caption{Solitary wave for $\gamma\!=\! 10$ and a representative value
of the velocity $v\!=\!c/2$. We show the radial dependence of the real
and the imaginary part of the wave function for various values
of $\xi$ in steps of $\delta\xi\!=\! 0.1$. The corresponding results 
for negative $\xi$ are obtained through the parity relations (\ref{eq:38}).
   }
   \label{fig:8}
\end{figure}

\begin{figure}
   \begin{center}
   \epsfig{file=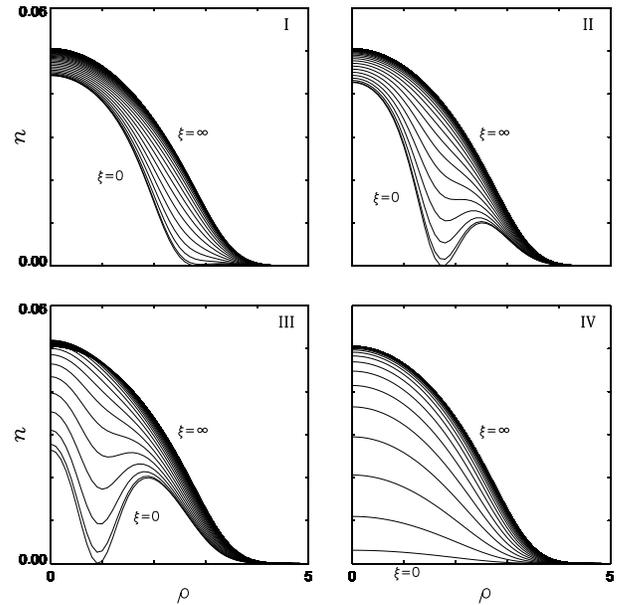,width=8cm,bbllx=35bp,bblly=190bp,bburx=545bp,bbury=700bp}
   \end{center}
   \caption{Radial dependence of the local particle density $n\!=\!|\Psi|^2$
for $\gamma\!=\! 10$, using the same conventions for the $\xi$ dependence 
as in Fig.~\ref{fig:8}. The four special cases considered correspond 
to the four representative points I, II, III, and IV along the
energy-momentum dispersion of Fig.~\ref{fig:7}.
   }
   \label{fig:9}
\end{figure}

\begin{figure}
   \begin{center}
   \epsfig{file=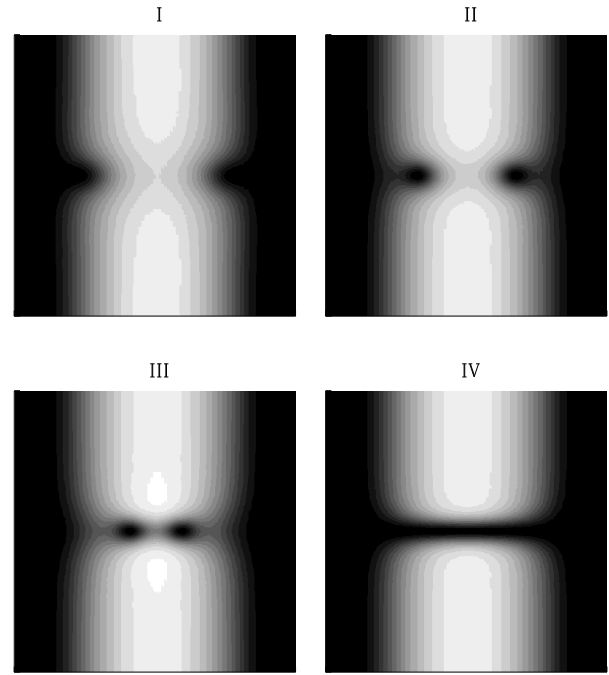,width=8cm,bbllx=25bp,bblly=140bp,bburx=565bp,bbury=740bp}
   \end{center}
   \caption{Contour levels of the local particle density $n\!=\!|\Psi|^2$
for $\gamma\!=\! 10$, in a plane that contains the $z$ axis and cuts across the
cylindrical trap. The complete 3D picture may be
envisaged by simple revolution around the $z$ axis.
Regions with high particle density are bright while regions with
zero density are black. The four special cases considered are the same as in
Fig.~\ref{fig:9}.
   }
   \label{fig:10}
\end{figure}

Hence we turn to a description of the detailed nature of this new 
sequence of solitary waves. For values of the velocity near the speed of 
sound $c$, the calculated soliton appears again as a mild soundlike
disturbance of the ground state. The dominant features of the solitary
wave are pronounced as the velocity is decreased to lower values and become
reasonably apparent for  $v\!=\!c/2$ that corresponds to point
I in the dispersion of Fig.\ref{fig:7}. The wave function
is completely illustrated through its real and imaginary parts in
Fig.~\ref{fig:8}.
An important new feature emerges by comparison with the corresponding
case at $\gamma\!=\! 1$ illustrated in frame I of Fig.~\ref{fig:4}.
Both the real and the imaginary part at the center of the soliton
($\xi\!=\!0$) now vanish for a specific radius $R\!=\! 2.8$,
thus a vortex ring is beginning to emerge. A partial but more transparent
illustration is given in Fig.~\ref{fig:9} where we depict the 
radial dependence of the local particle density $n\!=\! |\Psi|^2$
for various values of $\xi$. Again it is clear that the density near 
the center of the soliton
($\xi\!=\! 0$) vanishes on a ring with a relatively large radius
$R\!=\! 2.8\,$.
The features of the vortex ring become completely apparent,
and its radius is tightened, as we proceed to smaller values
of the velocity. A notable special case is the static ($v\!=\! 0$) vortex
ring with radius $R\!=\! 1.8$ illustrated in frame II of
Fig.~\ref{fig:9}, which is far from being a black soliton.
The corresponding point II in Fig.~\ref{fig:7} is thus a new local maximum
of the energy-momentum dispersion, which is clearly distinguished
from the local maximum at point IV that corresponds to the static
black soliton discussed earlier in the text.

One would think that pushing the velocity $v$ to negative values would 
somehow retrace the calculated sequence of vortex rings backwards.
In fact, our algorithm continues to converge to vortex rings of
smaller radii until the critical velocity $-v_1$ is encountered
where the ring achieves its minimum radius $R_{\rm min}\!=\! 0.8$ and
ceases to exist for smaller values of $v$. The terminal state at
$v\!=\!-v_1$ is illustrated in frame III of Fig.~\ref{fig:9}.
We have thus described a sequence of solitary waves that consists of
bonafide 3D vortex rings and does not contain a black soliton.
The corresponding branch in the energy-momentum dispersion
of Fig.~\ref{fig:7} is labeled by points I, II, and III
that stand for the special cases $v\!=\!c/2, 0$, and $-v_1$.
As mentioned already, an equivalent sequence of solitary waves exists in
the range $-c\!<\!v\!<\!v_1$ and leads to a dispersion curve in 
Fig.~\ref{fig:7} that is mirror symmetric to the branch (I,II,III) around
$Q\!=\!\pi$.

To complete the description for $\gamma\!=\! 10$ we must briefly return to 
the auxiliary sequence of solitary waves associated with the portion
of the dispersion that is depicted by a dotted line in Fig.~\ref{fig:7}.
As one moves from point III to point IV, the ringlike structure is more
or less preserved at constant radius $R\!=\!R_{\rm min}\!=\!0.8$.
Nevertheless, the detailed features of the vortex ring are tamed at small
velocities and completely disappear for $v\!=\!0$ to yield
a black soliton at point IV.

We thus essentially conclude our description of solitary waves for 
$\gamma\!=\! 10$ by schematically summarizing our main results in 
Fig.~\ref{fig:10}. Yet some of the elements of the preceding discussion
are sufficiently surprising to deserve closer attention.
For example, simple inspection of Fig.~\ref{fig:7} reveals that the
group velocity becomes negative in the region (II,III) or,
equivalently, the impulse is opposite to the group velocity.
This rotonlike behavior is consistent with the Onsager-Feynman view
of a roton as the ghost of a vanished vortex ring \cite{donnelly}
because the calculated radius of the vortex ring is monotonically
decreasing along the fundamental (I,II,III) sequence.
A full-scale roton would develop if the terminal point III were an
inflection point beyond which the group velocity begins to rise again.
Actually, this is exactly what happens as one moves away from point III
along the upper branch in Fig.~\ref{fig:7}, but this
``roton'' portion of the dispersion now appears in a strange location 
by comparison to the usual situation in liquid helium \cite{donnelly}.
On the other hand, the black soliton at the stationary point IV is
indeed the ghost of a vanished vortex ring, as explained in the preceding
paragraph.

\begin{figure}
   \begin{center}
   \epsfig{file=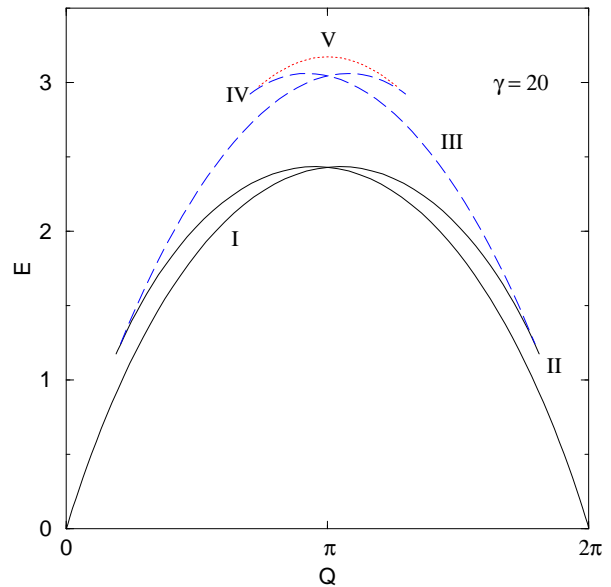,width=8cm}
   \end{center}
   \caption{Energy $E$ in units of $\gamma_\bot (\hbar\omega_\bot)$ versus
impulse $Q$ in units of $\gamma_\bot (\hbar/a_\bot)$ for $\gamma\!=\! 20$.
The solid lines correspond to the fundamental single-ring sequence,
the dashed lines to the double-ring sequence,
and the dotted line to the auxiliary sequence that contains
a black soliton (point V).
   }
   \label{fig:11}
\end{figure}

It is also interesting to question how the picture just described
evolves with increasing values of the dimensionless coupling constant $\gamma$
which is the only parameter that enters the rationalized GP equation.
Our numerical calculations have revealed yet another critical coupling
$\gamma_2\!\approx\! 12$, in the sense that new flavor arises for
$\gamma\!>\!\gamma_2$. The structure of the solitary waves in this
new regime becomes sufficiently clear for $\gamma\!=\! 20$ and
is best summarized by the calculated energy-momentum dispersion shown
in Fig.~\ref{fig:11}.
Apart from mirror symmetry, the dispersion now exhibits two cusps that
correspond to two critical velocities $v_1\!=\!1.35\!=\!0.64 c$ and 
$v_2\!=\!0.48\!=\! 0.23\, c$, where $c\!=\!2.1$ is the speed
of sound calculated for $\gamma\!=\!20$ as in Sec.~II.

\begin{figure}
   \begin{center}
   \epsfig{file=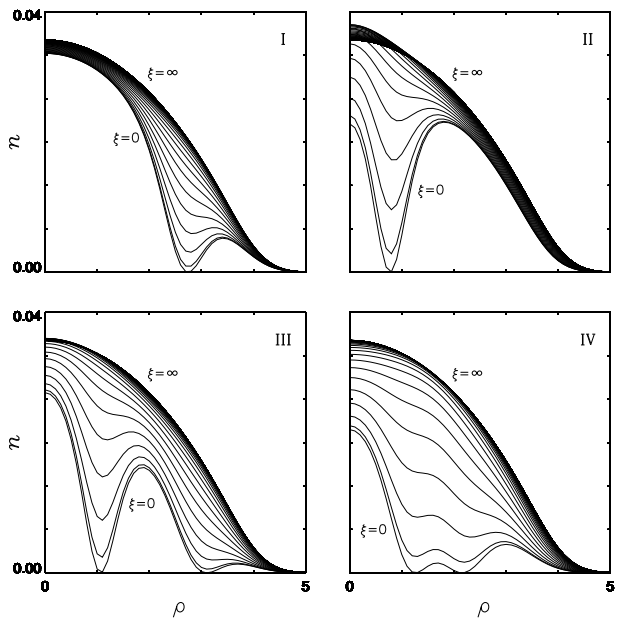,width=8cm,bbllx=40bp,bblly=190bp,bburx=545bp,bbury=700bp}
   \end{center}
   \caption{Radial dependence of the local particle density $n\!=\!|\Psi|^2$
for $\gamma\!=\! 20$. The four special cases considered correspond to
points I, II, III, and IV along the energy-momentum dispersion
of Fig.~\ref{fig:11}. Point V in the above dispersion in not illustrated here
because it corresponds to a black soliton similar to that shown earlier
in frame IV of Fig.~\ref{fig:9}.
   }
   \label{fig:12}
\end{figure}

The nature of the solitary waves associated with the various branches
in the dispersion of Fig.~\ref{fig:11} is very briefly described with
the aid of Fig.~\ref{fig:12}. Thus we consider the sequence of 
five characteristic points (I,II,...,V) that roughly cover half of
the dispersion, the other half being obtained by the mirror
symmetry (\ref{eq:39}). The lowest branch (I,II) corresponds
to single vortex rings with velocities in the range $c\!>\!v\!>\!-v_1$,
as in the case $\gamma\!=\!10$. Again the ring achieves its minimum
radius at the critical velocity $v\!=\!-v_1$ (point II).
The new element for $\gamma\!=\!20$ is the intermediate branch
(II,III,IV) that corresponds to double rings with velocities in the range
$v_2\!>\!v\!>\!-v_1$. The second ring is first created at the flanks
of the trap and comes closest to the original ring at the
new critical velocity $v\!=\!v_2$ (point IV). This double-ring
configuration is more or less preserved along the upper branch (IV,V),
with velocities in the range $v_2\!>\!v\!>\!0$, but gradually fades away
to become a black soliton at $v\!=\!0$ (point V).

While the numerical calculation becomes increasingly more difficult
for larger values of $\gamma$, it is clear that a sequence of 
critical couplings $\gamma_1,\gamma_2,\ldots$ exists and leads to
a hierarchy of axisymmetric vortex rings. The single-ring solution
associated with the lowest branch of the spectrum is a robust
feature for all $\gamma\!>\!\gamma_1\!\approx\!3.9$ and is likely stable 
to all perturbations. But it is possible that the multiple-ring configurations
associated with the higher branches are unstable to non-axisymmetric
perturbations. In this respect, one should recall that the solitary
wave that corresponds to the upper branch in the original calculation
of Jones and Roberts \cite{jones1} within the homogeneous GP model 
was later argued to be unstable \cite{jones2}.

However, we should emphasize that the vortex rings constructed here differ
significantly from the Jones-Roberts (JR) vortex ring that provided
the basic motivation for the present work. As with ordinary
smoke rings in fluid dynamics, the JR ring can never be static thanks
to a virial relation of the type (\ref{eq:34}) that prevents
finite-energy solutions with $v\!=\!0$ in the homogeneous GP model.
As a result, the radius of the vortex ring grows to infinity at low velocity.
This picture is completely rearranged in a cylindrical trap because
the occurrence of slow vortex rings with large radius is restricted by the
boundaries of the trap. Instead, vortex rings are predicted to nucleate at
the flanks of the trap as soundlike pulses with high velocity 
approaching the speed of sound $c$, and their radius actually decreases
with decreasing velocity. In particular, it is now possible
to obtain static ($v\!=\!0$) vortex rings of finite radius that are
no longer contradicted by the virial relation (\ref{eq:34}).
The structure of the energy-momentum dispersions calculated throughout
the present paper clearly reflects the substantial restructuring of  vortex
rings within a cylindrical trap.

It is then natural to question whether or not there exists a limit in which
the JR vortex ring is recovered. One should expect that this may
happen when the bulk healing length $L\!=\!(na)^{-1/2}$ is 
significantly smaller than the transverse oscillator length $a_\bot$.
This limit is translated into large values of $\gamma\!=\!\nu a$ which
is the only dimensionless parameter that enters the rationalized GP
equation. Now, our discussion earlier in this section suggests an
increasingly complicated hierarchical structure in the strong-coupling limit
rather that a simple JR soliton. A logical conclusion is that a JR vortex ring
somehow created within the bulk will sooner or later sense the
boundaries of the cylindrical trap. It will thus either directly
dissipate into sound waves, or reorganize itself to conform with one
or more of the presently calculated vortex rings possibly after 
ejecting some amount of radiation in the form of sound waves.

The preceding remarks indicate a certain non-uniformity that is
inherent in the approximation of the cigar-shaped trap by an infinitely-long
cylindrical trap. The same phenomenon is also apparent in the calculation
of linear modes in Sec.~II. For instance, neither one of the two
asymptotes for the speed of sound quoted in Eqs.~(\ref{eq:211}) and
(\ref{eq:212}) approaches the well-known Bogoliubov speed in a
homogeneous Bose gas \cite{jackson}. But a non-uniformity of this
type is not a reason to doubt that a sufficiently elongated trap
can be approximated by an infinitely-long cylindrical trap.

\section{Conclusion}
We have thus presented a complete 3D calculation of solitary waves in a
cylindrical Bose-Einstein condensate. In all cases considered there 
exists a nontrivial phase difference $\delta\phi$ that is reminiscent 
of strictly 1D solitons \cite{tsuzuki,zakharov} and is important
for their experimental production through phase imprinting
\cite{burger,denschlag}.

Nevertheless, the detailed structure of the solitary waves depends crucially
on the strength of the dimensionless effective coupling constant $\gamma$.
Quasi-1D solitons occur only in the weak-coupling region
$\gamma\!<\!\gamma_1\!\approx\!3.9$ where some of our accurate
numerical results could be approximated through effective 1D models
\cite{jackson2}. But a sufficiently strong coupling or density
is necessary in order to pronounce the special features of a
condensate. It is thus not surprising that the effective coupling in
experiments performed so far lies in the region $\gamma\!>\!\gamma_1$
where the nature of the theoretically predicted solitary waves
changes drastically.

For $\gamma\!>\!\gamma_1$ solitary waves are still characterized 
by a nontrivial phase difference $\delta\phi$ between the two ends of the
trap but are otherwise 3D vortex rings. This finding is consistent
with a recent experiment \cite{anderson} and a corresponding
theoretical analysis \cite{feder} in finite traps.
The main mathematical advantage of the approximation of a
sufficiently elongated trap by an infinitely-long cylindrical trap 
is that vortex rings can then be calculated in a steady state
propagating with a constant velocity $v$. It is thus possible to carry
out a detailed study of the soliton profile as a function of the effective
coupling constant $\gamma$ and the velocity $v$, as is done 
in the present paper.

An interesting by-product of the above idealization is that 
a soliton is characterized by a definite energy-momentum 
dispersion. The calculated dispersion is found to be the direct
analog of the Lieb mode \cite{lieb} in the weak-coupling region and
acquires interesting rotonlike features for stronger couplings. 
Perhaps such a dispersion can be measured by a combination of
phase imprinting \cite{burger,denschlag} with Bragg spectroscopy recently
employed for the detection of the usual Bogoliubov mode
\cite{stamper,ozeri}. In this respect, one should keep in mind that Bogoliubov
and Lieb modes operate at different energy and momentum scales in a
realistic trap. This fact becomes evident by the different sets
of physical units employed for the Bogoliubov mode in Fig.~\ref{fig:2}
and the Lieb mode in, say, Fig.~\ref{fig:6}. The difference is
accounted for by the second dimensionless coupling $\gamma_\bot\!=\!\nu a_\bot$
in Eq.~(\ref{eq:21}) which is much stronger than $\gamma\!=\!\nu a$
because $\gamma_\bot/\gamma\!=\!a_\bot/a \!\sim\! 10^2$.

\begin{acknowledgments}
We are grateful to A.R. Bishop, N.R. Cooper, G.M. Kavoulakis,
F.G. Mertens, and X. Zotos for valuable comments.
\end{acknowledgments}




\begin{thebibliography}{10}

\bibitem{dalfovo}  F. Dalfovo, S. Giorgini, L.P. Pitaevskii, and S. Stringari,
Rev. Mod. Phys. {\bf 71}, 463 (1999).

\bibitem{tsuzuki} T. Tsuzuki, J. Low Temp. Phys. {\bf 4}, 441 (1971).

\bibitem{zakharov} V.E. Zakharov and A.B. Sabat, Sov. Phys. JETP {\bf 37},
823 (1973).

\bibitem{kulish} P.P. Kulish, S.V. Manakov, and L.D. Faddeev,
Theor. Math. Phys. {\bf 28}, 615 (1976).

\bibitem{ishikawa} M. Ishikawa and H. Takayama, J. Phys. Soc. Japan
{\bf 49}, 1242 (1980).

\bibitem{lieb} E.H. Lieb, Phys. Rev. {\bf 130}, 1616 (1963).

\bibitem{lieb2} E.H. Lieb and W. Liniger, Phys. Rev. {\bf 130}, 1605 (1963).

\bibitem{burger} S. Burger, K. Bongs, S. Dettmer, W. Ertmer, K. Sengstock,
A. Sanpera, G.V. Shlyapnikov, and M. Lewenstein,
Phys. Rev. Lett. {\bf 83}, 5198 (1999).

\bibitem{denschlag} J. Denschlag, J. E. Simsarian, D. L. Feder, 
Charles W. Clark, L.A. Collins, J. Gubizolles, L. Deng, E.W.
   Hagley, K. Helmerson, W.P. Reinhardt, S.L. Rolston,
B.I. Schneider, and W.D. Phillips,
Science {\bf 287}, 97 (2000).

\bibitem{perez} V.M. P\'erez-Garcia, H. Michinel, and H. Herrero,
Phys. Rev. A {\bf 57}, 3837 (1998).

\bibitem{jackson} A.D. Jackson, G.M. Kavoulakis, and C.J. Pethick,
Phys. Rev. A {\bf 58}, 2417 (1998).

\bibitem{muryshev} A.E. Muryshev, H.B. van Linden van den Heuvell, and
G.V. Shlyapnikov, Phys. Rev. A {\bf 60}, R2665 (1999).

\bibitem{feder} D.L. Feder, M.S. Pindzola, L.A. Collins, B.I. Schneider,
and C.W. Clark, Phys. Rev. A {\bf 62}, 053606 (2000).

\bibitem{dikande} A.M. Dikande, e-print cond-mat/0111418 .

\bibitem{muryshev2} A.E. Muryshev, G.V. Shlyapnikov, W. Ertmer, K. Sengstock,
and M. Lewenstein, e-print cond-mat/0111506 .

\bibitem{salasnich} L. Salasnich, A. Parola, and L. Reatto,
e-print cond-mat/0201395 .

\bibitem{anderson} B.P. Anderson, P.C. Haljan, C.A. Regal, D.L. Feder,
L.A. Collins, C.W. Clark, and E.A. Cornell,
Phys. Rev. Lett. {\bf 86}, 2926 (2001).

\bibitem{jones1} C.A. Jones and P.H. Roberts, J. Phys. A: Math. Gen. {\bf 15},
2599 (1982).

\bibitem{semi}
N. Papanicolaou and P.N. Spathis, Nonlinearity {\bf 12}, 285 (1999).

\bibitem{komineas} S. Komineas and N. Papanicolaou, e-print cond-mat/0202182 .

\bibitem{dalfovo2} F. Dalfovo and S. Stringari, 
Phys. Rev. A {\bf 53}, 2477 (1996).

\bibitem{baym} G. Baym and C.J. Pethick, Phys. Rev. Lett. {\bf 76}, 6 (1996).

\bibitem{zaremba} E. Zaremba, Phys. Rev. A {\bf 57}, 518 (1998).

\bibitem{stringari} S. Stringari, Phys. Rev. A {\bf 58}, 2385 (1998).

\bibitem{fedichev} P.O. Fedichev and G.V. Shlyapnikov,
Phys. Rev. A {\bf 63}, 045601 (2001).

\bibitem{papspathis}
N. Papanicolaou and P.N. Spathis, J. Phys. G {\bf 11}, 149 (1985).

\bibitem{kavoulakis} G.M. Kavoulakis and C.J. Pethick, 
Phys. Rev. A {\bf 58}, 1563 (1998).

\bibitem{andrews} M.R. Andrews, D.M. Kurn, H.-J. Miesner, D.S. Durfee,
C.G. Townsend, S. Inouye, and W. Ketterle,
Phys. Rev. Lett. {\bf 79}, 553 (1997); {\it ibid} {\bf 80}, 2967 (1998).

\bibitem{batchelor} G.K. Batchelor, {\it An Introduction to Fluid Dynamics}
(Cambridge University Press, 1967).

\bibitem{saffman} P.G. Saffman, {\it Vortex Dynamics}
(Cambridge University Press, 1992).

\bibitem{jackson2} A.D. Jackson and G.M. Kavoulakis,
e-print cond-mat/0202178 .

\bibitem{donnelly} R.J. Donnelly, {\it Quantized Vortices in Helium} II
(Cambridge University Press, 1991).

\bibitem{jones2} C.A. Jones, S.J. Putterman, and P.H. Roberts, 
J. Phys. A: Math. Gen. {\bf 19}, 2991 (1986).

\bibitem{stamper} D.M. Stamper-Kurn, A.P. Chikkatur, A. G\"orlitz,
S. Inouye, S. Gupta, D.E. Pritchard, and W. Ketterle,
Phys. Rev. Lett. {\bf 83}, 2876 (1999).

\bibitem{ozeri} R. Ozeri, J. Steinhauer, N. Katz, and N. Davidson,
e-print cond-mat/0112496 .

\end{thebibliography}
\end{document}